\documentclass[reprint,prd,twocolumn,superscriptaddress,showpacs,showkeys,floatfix,preprintnumbers]{revtex4-2}
\usepackage{mathrsfs}
\usepackage{amsfonts}
\usepackage{amsmath}
\usepackage{slashed}
\usepackage{array}
\usepackage{verbatim}
\usepackage{epsfig}
\usepackage{graphicx}
\usepackage{color}
\usepackage{subfigure}
\usepackage{multirow}
\usepackage[dvipsnames]{xcolor}
\definecolor{ceruleanblue}{rgb}{0.16, 0.32, 0.75}
\usepackage[colorlinks=true,linkcolor=ceruleanblue,citecolor=ceruleanblue,urlcolor=blue]{hyperref}

\newcommand{\beq}{\begin{eqnarray}}
\newcommand{\eeq}{\end{eqnarray}}
\newcommand{\be}{\begin{equation}}
\newcommand{\ee}{\end{equation}}
\newcommand{\ba}{\begin{eqnarray}}
\newcommand{\ea}{\end{eqnarray}}

\newcommand{\bit}{\begin{itemize}}
\newcommand{\eit}{\end{itemize}}
\newcommand{\rw}{\rightarrow}

\newcommand{\zzuphy}{School of Physics, Zhengzhou University, Zhengzhou 450001, China}
\newcommand{\innovation}{Collaborative Innovation Center of Quantum Matter, Beijing 100871, China}
\newcommand{\chep}{Center for High Energy Physics, Peking University, Beijing 100871, China}
\newcommand{\pkuphy}{School of Physics, Peking University, Beijing 100871,
China}

\newcommand{\BNL}{Physics Department, Brookhaven National Laboratory, Upton, NY 11973, USA}

\begin{document}

\title{ Lattice calculation of the $\eta_c\eta_c$ and $J/\psi J/\psi$ s-wave scattering length}

\author{Yu Meng}
\email[Email: ]{yu\_meng@zzu.edu.cn}
\affiliation{\zzuphy}
\author{Chuan Liu}
\email[Email: ]{liuchuan@pku.edu.cn}
\affiliation{\pkuphy}\affiliation{\chep}\affiliation{\innovation}
\author{Xin-Yu Tuo}\affiliation{\BNL}
\author{Haobo Yan}\affiliation{\pkuphy}
\author{Zhaolong Zhang}\affiliation{\pkuphy}

\date{\today}

\begin{abstract}
We calculate the s-wave scattering length in the $0^+$ sector of $\eta_c\eta_c$ and the $2^+$ sector of
$J/\psi J/\psi$ using three $N_f=2$ twisted mass gauge ensembles
with the pion mass $300\text{-}365$ MeV and lattice spacing $a=0.0667,0.085,0.098$ fm, respectively. The scattering lengths are extracted using the conventional L{\"u}scher finite size method. We observe sizable discretization effects and the results after a continuum extrapolation are $a^{0^+}_{\eta_c\eta_c}=-0.104(09)$ fm and $a^{2^+}_{J/\psi J/\psi}=-0.165(16)$ fm. The slight difference of light quark mass is ignored in this study. Our results indicate that the interaction between the two respective charmonia are repulsive in nature in both cases.

\end{abstract}

\maketitle

\section{Introduction}
The last decades have witnessed continuing major discoveries of new resonant structures, many of which are also
called XYZ particles nowadays, due to their unknown nature. Typically, these structures are found to be
difficult to describe by the classification scheme of the conventional Quark Model(QM) because of their
generally nontrivial internal flavor structures which are confirmed by their decay products.
Ever since their discoveries, much attention from both experiments and
theories have been paid to the nature of these exotic states,
see Refs.~\cite{Chen:2016qju,Guo:2017jvc,Olsen:2017bmm,Brambilla:2019esw,Chen:2022asf} for recent reviews.
However, limited by insufficient knowledge of the highly nonperturbative properties
of Quantum chromodynamics(QCD) at low energies, the nature of most exotic states
remains obscure despite enormous progress made in phenomenological studies over the years.

In 2020, the LHCb Collaboration first reported an observation of a narrow structure around 6.9 GeV in the $J/\psi J/\psi$ invariance mass distribution with a global significance of more than $5\sigma$~\cite{LHCb:2020bwg}, called X(6900), which is supposed to be a promising candidate for a tetraquark $cc\bar{c}\bar{c}$ state consisting of pure charm flavor~\cite{Berezhnoy:2011xn,Wu:2016vtq,Wang:2017jtz,Anwar:2017toa,Karliner:2016zzc,Liu:2019zuc,Chen:2016jxd,Wang:2019rdo,Bedolla:2019zwg,Deng:2020iqw,Wang:2020ols,Jin:2020jfc,Lu:2020cns,Becchi:2020uvq,Sonnenschein:2020nwn,Giron:2020wpx,Karliner:2020dta,Karliner:2020dta,Zhao:2020nwy,Faustov:2020qfm,Zhang:2020xtb,Zhu:2020xni,Cao:2020gul,Gong:2020bmg,Yang:2020wkh,Huang:2020dci,Zhao:2020jvl,Ke:2021iyh,Mutuk:2021hmi,Li:2021ygk,Wang:2021kfv,Tiwari:2021tmz,Liu:2021rtn,Kuang:2022vdy,Wu:2022qwd,Ortega:2023pmr,liu:2020eha,Chen:2024bpz,Wu:2024euj,Wu:2024hrv}.
Most recently, ATLAS and CMS collaborations confirmed this exotic state and found more broad structures~\cite{ATLAS:2023bft,CMS:2023owd}.
Moreover, a near-threshold state called X(6200) in the $J/\psi J/\psi$ system
with the quantum number $J^{PC}=0^{++}$ or $2^{++}$ is suggested theoretically
through a coupled-channel analysis~\cite{Dong:2020nwy,Song:2024ykq} of the LHCb data on the $J/\psi J/\psi$ spectrum,
more studies can be found in Ref.~\cite{Dong:2021lkh,Liang:2021fzr,Nefediev:2021pww}.
So far, most of the theoretical studies on the fully-charm tetraquarks states
are quite model-dependent, and the conclusions on the nature of the state
can be quite different even though many studies can describe
the experimental data reasonably well. Hence, further experimental and model-independent theoretical
studies are pivotal and timely in providing more information on various possible interpretations. It will lead to a better understanding of the nature of these exotic states.
As a genuine nonperturbative method, lattice QCD is supposed to be an ideal tool to study
these multi-quark systems and it is also expected to provide new insights into the QCD low-energy behaviors.

In the past decades, the precision of lattice calculation has been greatly improved,
even up to the sub-percent level, and with all systematic effects included in the strong interaction
physics~\cite{FlavourLatticeAveragingGroupFLAG:2021npn}. However, due to various technical and unique difficulties
in a fully-charmed tetraquark system, for example, the multi-channel nature and
the mixing of various single- and multi-particle states,
a systematic lattice study remains difficult to date. It continues to be a booming field in the future.
For the reasons given above, it is interesting and meaningful to study a simple but general situation nowadays.

\begin{figure}[!h]
\centering
\subfigure{\includegraphics[width=0.34\textwidth]{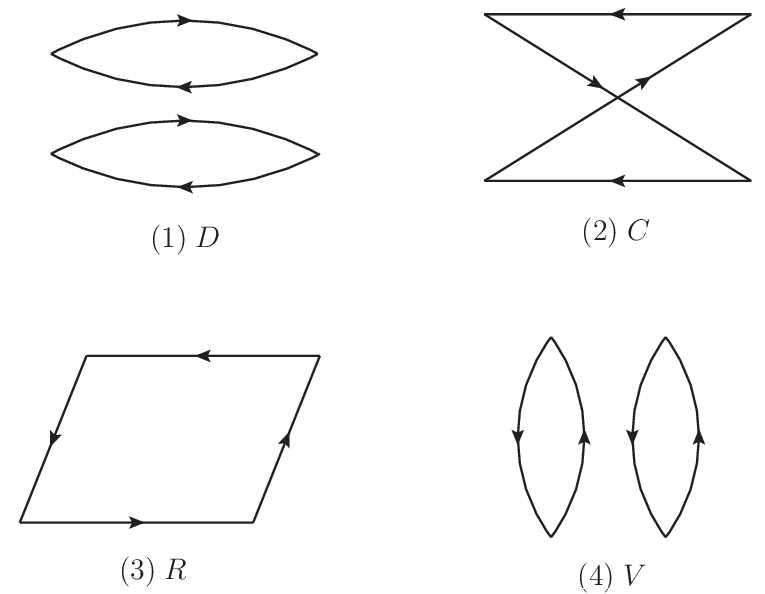}}\hspace{5mm}
\caption{All diagram contributions to $\eta_c\eta_c$ or $J/\psi J/\psi$ four-point functions, which are denoted
by the direct (D), crossed (C), rectangular (R), and vacuum (V), respectively. In this work, only (D) and (C) 
are considered.}
\label{fig:diagram}
\end{figure}

In this work, we present a single-channel lattice calculation on the s-wave scattering length
of $\eta_c\eta_c$ system with the quantum number $J^{PC}=0^{++}$
and $ J/\psi J/\psi$ system with $J^{PC}=2^{++}$ using the standard
L{\"u}scher finite size method~\cite{Luscher:1985dn,Luscher:1986pf,Luscher:1990ck,Luscher:1990ux}.
The method establishes a relationship between the energies of a two-particle system
in a finite volume and the scattering phase in the infinite volume,
thus providing a direct way to extract the scattering length from the lattice simulation.
In Fig.~\ref{fig:diagram}, we display the possible quark line diagrams contributing
to the $\eta_c\eta_c$ and $J/\psi J/\psi$ four-point functions.
For the charmonium scattering, the diagrams of the type (R) and (V) suffer from
a suppression by the strong coupling constant at the charm quark scale. Therefore, in this exploratory calculation, these two types are neglected and we only consider the contributions of types (D) and (C) in Fig.~\ref{fig:diagram}.

The rest of this paper is organized as follows. In Sec.~\ref{sec:method}, we
introduce the  L{\"u}scher method utilized in this work to extract the scattering length. In Sec.~\ref{sec:setup}, the configuration information is given. In Sec.~\ref{sec:result}, we give details of the simulations and show the main results. Finally, we conclude in Sec.~\ref{sec:conclude}.

\section{Methodology}\label{sec:method}
We consider the scattering process of $\eta_c\eta_c$ and $J/\psi J/\psi$ in the rest frame.
Particularly, we focus on the channel of $J^{PC}=0^{++}$ for $\eta_c\eta_c$
and $J^{PC}=2^{++}$ for $J/\psi J/\psi$, respectively.
Both of the scattering lengths could be obtained by the L{\"u}scher method, which relates
the energy levels in a finite volume to the scattering phase in an infinite volume.
As for the system of two identical vector particles, possible combinations of total spin $\mathcal{S}$
and the angular momentum $\mathcal{L}$ must satisfy the Bose symmetry and is a bit involved
than the $\eta_c\eta_c$ case. However, as far as the $2^{++}$ state is concerned,
there are four possible combinations
$\{\mathcal{S},\mathcal{L}\}=[\{2,0\},\{0,2\},\{2,2\},\{2,4\}]$.
A coupled-channel formula in this state has been established in Ref.~\cite{Romero-Lopez:2018zyy}.
If one ignores the mixing of the g-wave, the four-channel L{\"u}scher equation then reduces
to a single-channel equation,
which is equivalent to describing the system using
the single channel quantum number $\{\mathcal{S},\mathcal{L}\}=\{2,0\}$.

In the rest frame, the lattice has a cubic symmetry whose irreducible representations (irreps) $\Gamma$
is to be related to certain angular momentum $J$ in the infinite volume. For example,
the $0^+$ state in an infinite volume corresponds to $A_1$ irrep in the finite volume, and $2^+$ is related to the $E,T_2$ irreps.
Using the L{\"u}scher finite size method, one can extract the corresponding scattering length
$a_{\eta_c\eta_c}^{0^+}$ and $a_{J/\psi J/\psi}^{2^+}$ from the lowest energy shift
\be
\label{eq:energy_shift}
\delta E^{\Gamma}=-\frac{4\pi a^{\Gamma}}{mL^3}\left[1+c_1\frac{a^{\Gamma}}{L}+
c_2\left( \frac{a^{\Gamma}}{L}\right)^2+\mathcal{O}(L^{-3}) \right]
\ee
where the symbol $\Gamma$ denotes the irrep $A_1$ for $0^{+}$ channel in $\eta_c\eta_c$ and
$E$ or $T_2$ for $2^+$ channel in $J/\psi J/\psi$. For convenience, we introduce $a^{A_1}$
to denote $a_{\eta_c\eta_c}^{0^+}$ and $a^{E/T_2}$ for $a_{J/\psi J/\psi}^{2^+}$.
The purely numerical constants $c_1=-2.837297$, $c_2=6.375183$, and $L$ represent the size of the cubic volume.
Note the mass $m=m_{\eta_c}$ for $\Gamma=A_1$ and $m=m_{J/\psi}$ for $\Gamma=E,T_2$.

To extract $\delta E^{\Gamma}$, we first construct the two-point functions for $\eta_c$ and $J/\psi$:
\be
\left\{\begin{aligned}
C_{\eta_c}^{(2)}(t)&=\frac{1}{T}\sum\limits_{t_s}\langle \mathcal{P}(t+t_s)\mathcal{P}^{\dagger}(t_s) \rangle\;,
\\
C_{J/\psi}^{(2)}(t) &=\frac{1}{3T}\sum\limits_{t_s}\sum\limits_{i=1,2,3}\langle \mathcal{V}_i(t+t_s)\mathcal{V}_i^{\dagger}(t_s) \rangle\;.
\end{aligned}\right.
\ee
where $\mathcal{P}(t)\equiv \bar{c}\gamma_5c(t)$ and $\mathcal{V}_i(t)=\bar{c}\gamma_ic(t)$ represents
the interpolating operator of $\eta_c$ and $J/\psi$, respectively. The ground state energies are
then extracted from a two-state fit of the two-point functions computed in the lattice simulation:
\be
\label{eq:2pt}
C^{(2)}_h(t)=V \sum_{i=0,1}\frac{(Z_i^h)^2}{2E_i^h} \left(e^{-E_i^ht}+e^{-E_i^h(T-t)}\right)
\;,
\ee
where the symbol $h$ denotes the $\eta_c$ or $J/\psi$ particle, $V$ is the spatial-volume, $E_0^{\eta_c}=m_{\eta_c}$, $E_0^{J/\psi}=m_{J/\psi}$ denoting the ground-state energy and
$E_1^h$ is the energy of the first excited state. The factors $Z_i^{\eta_c}=\frac{1}{\sqrt{V}}\langle i|\mathcal{P}^\dagger(0)|0\rangle$, $Z_i^{J/\psi}=\frac{1}{\sqrt{V}}\langle i|\mathcal{V}^\dagger(0)|0\rangle$ ($i=0,1$) are the overlap amplitudes for the ground and the first excited states.

Next, we construct the four-point functions for these two charmonia systems,
$ \mathcal{O}^{A_1}(t)$ for $\eta_c\eta_c$ and $\mathcal{O}^{E}(t)$, $\mathcal{O}^{T_2}(t)$
for $J/\psi J/\psi$ system~\cite{Romero-Lopez:2018zyy}
\beq
\mathcal{O}^{A_1}(t) &=&\mathcal{P}(t)\mathcal{P}(t)\;,
\\
\mathcal{O}^{E}(t)&=&\Big{\{} \frac{1}{\sqrt{2}}\left[\mathcal{V}_1(t)\mathcal{V}_1(t)-\mathcal{V}_2(t)\mathcal{V}_2(t)\right],
\\
&& \frac{1}{\sqrt{2}}\left[\mathcal{V}_2(t)\mathcal{V}_2(t)-\mathcal{V}_3(t)\mathcal{V}_3(t)\right] \Big{\}}
\nonumber \\
\mathcal{O}^{T_2}(t)&=&\Big{\{} \mathcal{V}_2(t)\mathcal{V}_3(t) , \mathcal{V}_3(t)\mathcal{V}_1(t) ,
\mathcal{V}_1(t)\mathcal{V}_2(t) \Big{\}}
\eeq
In each symmetry channel,
the above two charmonia operators $\mathcal{O}^{\Gamma}(t_s)$
are then placed at a given time-slice $t_s$ and correlated with
another operator at time-slice $t+t_s$. To enhance the signal, we measured all time-slices
and the final four-point function is the average over different time slices:
\beq
\label{eq:four_point}
C^{(4)}_{\Gamma}(t)=\frac{1}{T}\sum\limits_{t_s}\langle \mathcal{O}^{\Gamma}(t+t_s)
\left(\mathcal{O}^{\Gamma}(t_s)\right)^{\dagger}\rangle
\eeq
To further remove possible constant terms in the four-point function,
we construct the following ratio~\cite{Feng:2009ij}
\be\label{eq:ratio}
R^{\Gamma}(t)=\frac{ C_{\Gamma}^{(4)}(t)-C_{\Gamma}^{(4)}(t+1)}{\left(C_h^{(2)}(t)\right)^2-
\left(C_h^{(2)}(t+1)\right)^2}
\ee
where $h=\eta_c$ for $\Gamma=A_1$ and $h=J/\psi$ for $\Gamma=E,T_2$.
The ratio has the advantage that it directly contains the parameter $\delta E^\Gamma$
that enters L\"uscher's formula~(\ref{eq:energy_shift}) for the scattering length.
For a large time separation $t \gg 1$, the ratio has the following asymptotic form
\be
\label{eq:ratio_final}
\!\!\!\!\!\!\!R^{\Gamma}(t)
= A_R[\cosh(\delta E^{\Gamma}t')+\sinh(\delta E^{\Gamma}t')\coth(2m_ht')],
\ee
where $t'=t+1/2-T/2$ and $A_R$ is an unknown coefficient.
The hadronic mass $m_h$ in this formula is extracted from the two-state fit
from Eq.~(\ref{eq:2pt}). Thus, the energy shift $\delta E^{\Gamma}$ can be
extracted straightforwardly by fitting the simulation data of $R^{\Gamma}(t)$
to the above equation.

\section{Lattice setups}\label{sec:setup}
\begin{table}[!h]
\caption{
Parameters of gauge ensembles are used in this work. From left to right, we list the ensemble name, the lattice spacing $a$, the spatial and temporal lattice size $L$ and $T$, the number of the measurements
of the correlation function for each ensemble $N_{\textrm{conf}}\times T$ with $N_{\textrm{conf}}$ the number of the configurations used, the pion mass $m_{\pi}$, and the spatial lattice size $L$ in physical unit.}
\label{tab:cfgs}
\begin{ruledtabular}
\begin{tabular}{cccccc}
\textrm{Ensemble} & $a$ [fm]  &$L^3\times T$ & $N_{\textrm{conf}}\times T$
& $m_{\pi} [\textrm{MeV}]$ & $L$[fm] \\
\hline
a67 & 0.0667(20)  & $32^3\times 64$& $197\times 64$ & 300 & 2.13 \\
a85 & 0.085(2)  & $24^3\times 48$ & $200\times 48$ & 315 & 2.04\\
a98 & 0.098(3)  & $24^3\times 48$ & $236\times 48$ & 365 & 2.35\\
\end{tabular}
\end{ruledtabular}
\end{table}

In this work, we use three two-flavor twisted mass gauge ensembles generated by
the Extended Twisted Mass Collaboration (ETMC)~\cite{ETM:2009ptp,Becirevic:2012dc} with lattice spacing
$a \simeq 0.0667,0.085,0.098$ fm. We call these ensembles a67, a85, and a98, respectively.
The ensemble parameters are shown in Table.~\ref{tab:cfgs}. The valence charm quark mass is tuned
by setting the lattice result of $J/\psi$ mass to the physical one.
Further detailed information can be found in Ref.~\cite{Meng:2021ecs}.

We adopt the $Z_4$-stochastic wall-source to compute the two-point function $C_h^{(2)}(t)$
and four-point function $C_{\Gamma}^{(4)}(t)$. All propagators are produced on all time slices so that we
can average over them to increase the statistics based on time translation invariance. We also apply the APE~\cite{APE:1987ehd} and Gaussian smearing~\cite{Gusken:1989qx} to the quark field to efficiently reduce the excited-state effects. For all the ensembles used in this work, we find an obvious plateau at the distance $t \gtrsim 1.7$ fm for the effective energies extracted from either $C_h^{(2)}(t)$ or $C_{\Gamma}^{(4)}(t)$.
As the excited-state effects in $C_{\Gamma}^{(4)}(t)$ and $C_h^{(2)}(t)$ partly cancel each other,
the ratio $R^{\Gamma}(t)$ exhibits a ground-state dominance from about $t\simeq 1.4$ fm.

\section{Lattice results}\label{sec:result}

\begin{table}[!h]
\center
\caption{ Numerical results of energy shift $\delta E^{\Gamma}$ in physical units and scattering length $m_{J/\psi}a^{\Gamma}$ for all ensembles. The errors of $\delta E^{\Gamma}$ do not include the 
lattice spacing error. $a^{\Gamma}$ is the scattering length of the channel
$\Gamma=A,E,T_2$. The continuous extrapolation for the dimensionless quantity $m_{J/\psi} a^{\Gamma}$ is performed
in $a^2\rw 0$, and the physical scattering length $a^{\Gamma}$ is then obtained after taking into account the physical value $m_{J/\psi}=3.0969$ GeV.}
\label{tab:results}
\begin{ruledtabular}
\begin{tabular}{ccccc}
Ensemble & $\Gamma$ & $A_1$ & $E$ &$T_2$ \\
\hline
a98 &	$\delta E^{\Gamma}[\textrm{MeV}]$ & 0.59(07) & 1.07(17) & 1.18(14) \\
a85 &	$\delta E^{\Gamma}[\textrm{MeV}]$ & 1.40(11) & 2.43(25) & 2.39(20) \\
a67 &	$\delta E^{\Gamma}[\textrm{MeV}]$ & 1.42(07) & 2.50(20) & 2.57(15) \\
\hline
a98 & $m_{J/\psi}a^{\Gamma}$ & $-$0.705(81) & $-$1.22(17) & $-$1.34(14) \\
a85 & $m_{J/\psi}a^{\Gamma}$ & $-$1.042(72) & $-$1.70(15) & $-$1.68(12) \\
a67 & $m_{J/\psi}a^{\Gamma}$ &$-$1.202(51) & $-$1.97(13) & $-$2.01(10) \\
Cont.Limit& $m_{J/\psi}a^{\Gamma}$ &$-$1.63(14) & $-$2.63(31) & $-$2.60(25) \\
\hline
Cont.Limit& $a^{\Gamma}[\textrm{fm}]$ &$-$0.104(09) & $-$0.168(20) & $-$0.165(16) \\
\end{tabular}
\end{ruledtabular}
\end{table}

\begin{figure}[!h]
\centering
\subfigure{\includegraphics[width=0.48\textwidth]{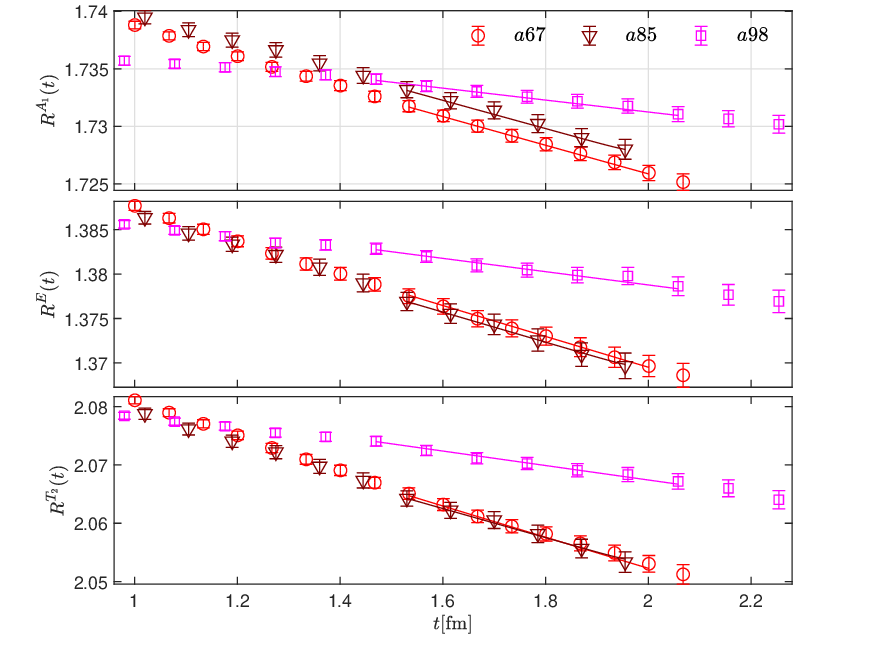}}\hspace{5mm}
\caption{Lattice results of $m_{J/\psi}a^{\Gamma}$ as a function of $t$ for all ensembles, $\Gamma=A_1,E,T_2$
from top to bottom.}
\label{fig:R}
\end{figure}

In Fig.~\ref{fig:R} we show the lattice results of $R^{\Gamma}(t)$ as a function of time $t$, with
$\Gamma=A_1,E,T_2$ from the top panel to the bottom. For all ensembles, we perform a correlated fit of the lattice data to the asymptotic form given in Eq.~(\ref{eq:ratio}) at a unified time window $t \sim [1.5,2.0]$ fm, and
extract the energy shift $a\delta E^{\Gamma}$. Using the l{\"u}scher method, we immediately obtain the dimensionless
scattering length $m_{J/\psi}a^{\Gamma}$. Both the results of $a\delta E^{\Gamma}$ and $m_{J/\psi}a^{\Gamma}$ are summarized in Table.~\ref{tab:results}.

\begin{figure}[!h]
\centering
\subfigure{\includegraphics[width=0.48\textwidth]{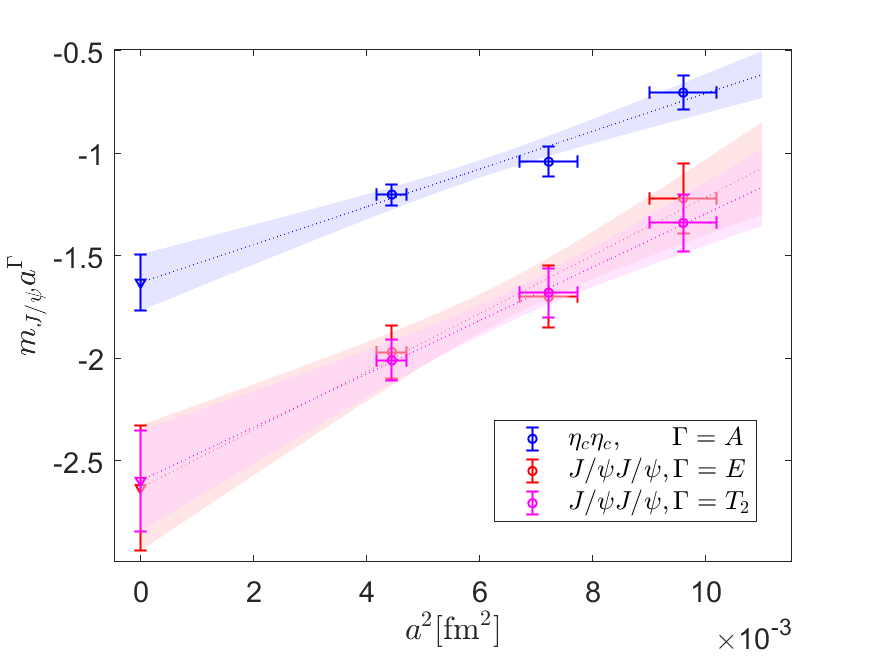}}\hspace{5mm}
\caption{Lattice results of $m_{J/\psi}a^{\Gamma}$ as a function of lattice spacing squared.
The errors of lattice spacing are included in the fitting and presented by the horizontal error bars.
The symbols of the circle denote the lattice results of ensemble a67,a85, and a98 from left to right.
The triangles are the results in the continuum limit $a^2\rw 0$.}
\label{fig:cont_limit}
\end{figure}

In Fig.~\ref{fig:cont_limit}, an extrapolation of $m_{J/\psi}a^{\Gamma}$ linear in $a^2$ is performed using three
different lattice spacings. Such a behavior is expected for the twisted mass configuration which has the so-called
automatic $O(a)$ improvement. It is seen that the linear fitting curve adequately describes all of
our lattice results nicely, indicating that no ensemble we utilized has a residual $\mathcal{O}(a)$ effect.
The same conclusion has also been demonstrated and confirmed in previous lattice studies
~\cite{ETM:2009ptp,Alexandrou:2009qu,ETM:2009ztk,Becirevic:2012dc,Meng:2021ecs,Zou:2021mgf,Meng:2024axn}.
Besides, the valence light quark is only involved in the sea for the charmonium system, the slightly different light quark masses should have little effects and we therefore ingore the difference of the light quark mass in this work. After a continuum extrapolation for the dimensionless $m_{J/\psi}a^{\Gamma}$, we rescale them to
physical values by the experimental mass of $J/\psi$, \textit{i.e.} $m_{J/\psi}^{\textrm{exp}}=3.0969$ GeV.
The physical scattering length is then obtained and tabulated
in Table.~\ref{tab:results}, where the errors have already included the uncertainty
of lattice spacing.

In the continuum limit, it is found that $a^{E}$ and $a^{T_2}$ are completely consistent with each other,
which is expected since both of them describe the same $2^{+}$ state of $J/\psi$-$J/\psi$ system.
In this paper, we take the result of $\Gamma=T_2$ with minor uncertainty as our final estimation
of the $2^+$ state. So, we present the scattering length of $0^+$ $\eta_c\eta_c$ and
$2^+$ $J/\psi J/\psi$ as
\beq
\begin{split}
a_{\eta_c\eta_c}^{0^+}&=&-0.104(09) \; \textrm{fm} \\
a_{J/\psi J/\psi}^{2^+}&=&-0.165(16) \; \textrm{fm}
\end{split}
\eeq
where the errors are only statistical and the lattice spacing errors are already included in the continuum limit.

\section{Discussion}
A recent phenomenological study suggests that the existence of a $J/\psi J/\psi$ bound state is plausible through the exchange of soft gluons~\cite{Dong:2021lkh}. Concurrently, interactions between pairs of bottomonia \cite{Brambilla:2015rqa} or $\Omega_{ccc}$ \cite{Lyu:2021qsh} are found to be attractive and may even be strong enough to produce a near-threshold bound state. In the cases discussed above, the interactions between two hadrons or baryons are predominantly governed by soft gluon exchange. At relatively long distances, gluons can hadronize into light-quark hadrons, such as pions and kaons, among others. The soft gluon contribution primarily originates from the direct diagram of type (D) as shown in Figure \ref{fig:diagram}. The crossed diagram of type (C) is, in fact, associated with $c\bar{c}$ exchange. In previous lattice calculations \cite{Lyu:2021qsh} and phenomenological studies \cite{Dong:2021lkh,Brambilla:2015rqa}, these crossed contributions were either non-existent or overlooked. Therefore, it is essential to investigate the separate contribution of type (D) to determine if its interaction is indeed attractive.

\begin{table}[!h]
\center
\caption{Numerical results of energy shift $\delta E$ from the type (D) and type (C), respectively. The total contribution of (D)+(C) is also presented for a better comparison. All results are in physical units.}
\label{tab:seperate_results}
\begin{ruledtabular}
\begin{tabular}{ccccc}
Ensemble  & $\delta E[\textrm{MeV}]$ & $A_1$ & $E$ &$T_2$ \\
\hline
\multirow{3}{*}{a98} &	(D) & $-$0.63(07) & $-$1.17(16) & $-$1.06(12) \\	
& (C) & $-$7.59(16) & $-$5.82(24) & $-$5.72(24) \\
& (D)+(C) & 0.59(07) & 1.07(17) & 1.18(14) \\
\hline
\multirow{3}{*}{a85} &	(D)& $-$1.12(11) & $-$1.85(24) & $-$1.73(18) \\
 &	 (C) & $-$14.81(22) & $-$10.36(41) & $-$9.97(38) \\
 &	(D)+(C)  & 1.40(11) & 2.43(25) & 2.39(20) \\
\hline
\multirow{3}{*}{a67} & (D) & $-$1.10(07) & $-$2.14(17) & $-$2.08(13) \\
 &	 (C) & $-$15.38(13) & $-$11.57(28) & $-$11.39(27) \\
 &	(D)+(C) & 1.42(07) & 2.50(20) & 2.57(15) \\
\end{tabular}
\end{ruledtabular}
\end{table}

\begin{figure}[!h]
\centering
\subfigure{\includegraphics[width=0.48\textwidth]{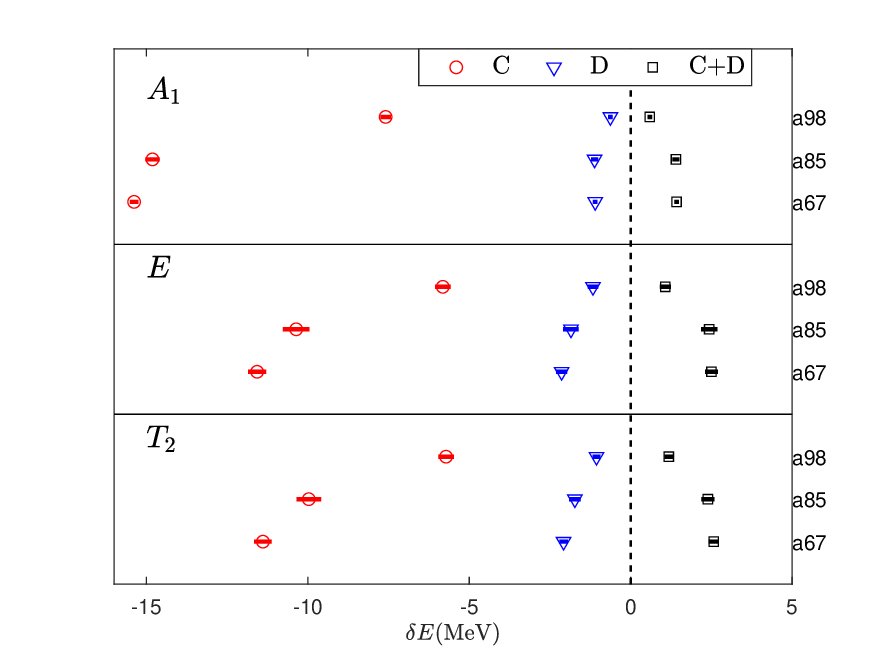}}\hspace{5mm}
\caption{A summary of energy shift $\delta E$ in $A_1,E,T_2$ channels from the different quark diagrams.
The red circles denote the contribution from the type (C). The blue triangles come from the type (D). The
total results of (C)+(D) are described by the black boxes. The numerical values of $\delta E$ are listed in the Tab.~\ref{tab:seperate_results}.}
\label{fig:dE_D_C}
\end{figure}

The energy shift $\delta E$ corresponding to individual diagrams can be extracted directly. For all ensembles, the results of $\delta E$ in the $A_1,E,T_2$ channels are depicted in Fig.~\ref{fig:dE_D_C}, and the numerical values are compiled in Tab.~\ref{tab:seperate_results}. The contribution from the soft gluon exchange, \textit{i.e.}, type (D), is found to be attractive. The energy shifts for all ensembles are approximately 1\text{-}2 MeV, which is in good agreement with the phenomenological estimation \cite{Dong:2021lkh}. The crossed diagram not only results in a relatively large negative energy shift individually but also alters the nature of the interaction when combined with the soft gluon contribution. Hence, the crossed diagram of type (C) plays a pivotal role and should not be ignored in the double-charmonium scattering process. More detailed discussions are collected in the Appendix.~\ref{sec:appendix}. The first observation of such phenomenon can be traced back to a lattice nonrelativistic QCD study on the $\bar{b}\bar{b}bb$ system \cite{Hughes:2017xie}, where the authors found no evidence of a $\bar{b}\bar{b}bb$ tetraquark with a mass below the lowest thresholds in the $0^+$ and $2^+$ channels, even though each diagram, \textit{e.g.}, type (D) or type (C), contributes to an attractive interaction. More importantly, in a real lattice calculation, it is only when all quark diagrams are summed that the physical result can be obtained. Therefore, we conclude that there is no indication of an attractive interaction in any channel studied in this work. 

Note that the light quark masses used in our calculations remain nonphysical. Since the direct diagram of type (D) is primarily dominated by the exchange of soft gluons, the strength of the attraction may be influenced by the light quark masses~\cite{Fujii:1999xn}. Consequently, a further lattice study employing a physical pion mass is warranted. Another significant systematic effect not considered in this work is the total contribution from type (R) and (V) diagrams, as shown in Fig.~\ref{fig:diagram}. These contributions have been accounted for in the calculations of the $I=0$ $\pi\pi$ scattering process~\cite{Liu:2016cba,Briceno:2016mjc,Fu:2017apw,RBC:2021acc}. Although the contribution from type (R) is as important as type (C) in $I=0$ $\pi\pi$ scattering, the scenario in double-charmonium scattering differs significantly. Firstly, type (V) in the $J/\psi J/\psi$ channel is closely associated with the coupling $\alpha_s^3(2m_c)$, while type (V) in the $\eta_c\eta_c$ channel is related to $\alpha_s^2(2m_c)$, where $\alpha_s(2m_c)$ denotes the strong coupling constant at the scale of $2m_c$. Secondly, the coupling for type (R) in both channels should be $\alpha_s(2m_c)$, considering that the intermediate single-hadron contains charmonium components $c\bar{c}$, and the energy carried by the gluon is nearly $2m_c$. Because the charm quark mass is substantially greater than the light quark mass, it follows that $\alpha_s(2m_c) \ll \alpha_s(2m_l)$. Ultimately, these lead to a significant suppression of types (R) and (V) in the double-charmonium scattering process. Direct computation of diagrams (R) and (V) in the charmonium sector remains challenging due to typical intermediate state contamination in the computation of the four-point function $C^{(4)}_{\Gamma}(t)$, where all possible single-particle states could mix, leading to severe signal-to-noise problems. Nonetheless, these intermediate contributions are all suppressed by the strong coupling constant at the charm quark scale, and thus should have a minimal contribution to the double-charmonium scattering process.

\section{Conclusion}\label{sec:conclude}
In this work, we present the first-principle calculation on the s-wave scattering length of a double charmonia system using three $N_f=2$ twisted mass gauge ensembles with pion masses ranging from 300 MeV to 365 MeV and  lattice spacing $a=0.0667,0.085,0.098$ fm, respectively.
In particular, the s-wave scattering length of $0^+$ $\eta_c\eta_c$ and $2^+$ $J/\psi J/\psi$ are obtained using the L{\"u}scher finite size method. A significant discretization effect is observed and eliminated by a continuum extrapolation. The light quark mass dependence is expected to be small for a pure charmonium system, therefore the difference of light quark masses in three ensembles are ignored in this work. Finally, we obtain the results as $a^{0^+}_{\eta_c\eta_c}=-0.104(09)$ fm and
$a^{2^+}_{J/\psi J/\psi}=-0.165(16)$ fm, with the uncertainties of lattice spacing
already taken into account in the final errors. 

The individual quark diagrams for soft gluon exchange and the $c\bar{c}$ exchange have been investigated, and both are found to contribute to an attractive interaction. The contribution from the crossed diagram is crucial and should not be disregarded in this process. When these two contributions are summed up together, all channels indicate repulsive interactions, with no evidence of possible bound states in the $0^+$ $\eta_c \eta_c$ and $2^+$ $J/\psi J/\psi$ channels. In our calculations, certain parts of the quark propagation diagrams, such as the rectangular and vacuum diagrams, as well as the high partial-wave mixing in the $2^+$ $J/\psi J/\psi$ channel, are anticipated to have minimal impact and are thus negligible in the current calculations. The light quark masses used in our calculations are still nonphysical. These effects must be taken into account in future more comprehensive investigations.

\begin{acknowledgments}
We thank ETM Collaboration for sharing the gauge configurations with us. Y.M. is grateful to Xu Feng and Feng-Kun Guo for helpful discussions and comments on reading through the manuscript. The authors also thank Dmitri Kharzeev and Nora Brambilla for useful comments on the preprint manuscript.
The authors acknowledge support by NSFC of China under Grant 
No. 12305094, 12293060, 12293061, 12293063, 11935017. The calculation was carried out on the Tianhe-1A supercomputer at Tianjin National
Supercomputing Center and the SongShan supercomputer at the National Supercomputing Center in Zhengzhou.

\end{acknowledgments}

\appendix
\section{Contributions from different diagrams}\label{sec:appendix}
\begin{figure*}[!htb]
\resizebox{0.45\textwidth}{!}{\includegraphics{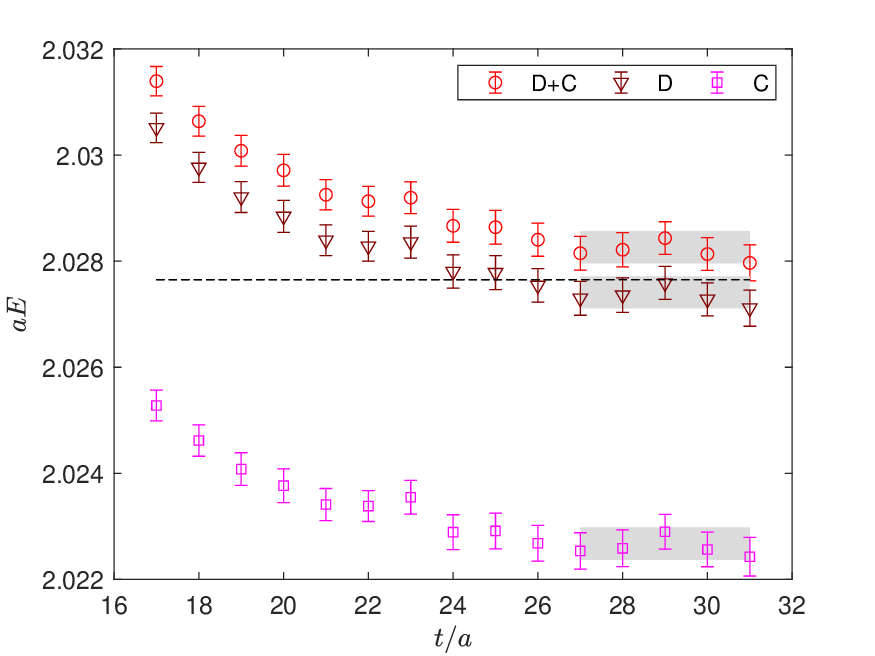}}
\resizebox{0.45\textwidth}{!}{\includegraphics{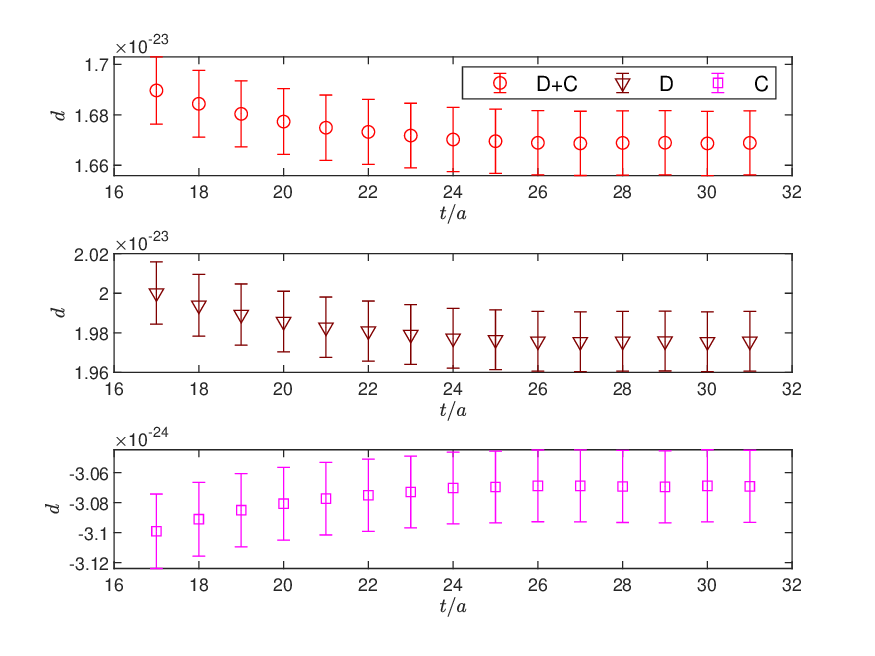}}
\caption{For the $A_1$ channel of ensemble a67, the two-particle energies $aE$ and coefficients $d$ for different diagrams D+C, D, and C, respectively. The dashed black line is the free energy of double-$\eta_c$ particles. $E$ and $d$ are introduced in Eq.~(\ref{eq:c4_mass2}).}
\label{fig:A1_mass2}
\end{figure*}

In this appendix, we take the $A_1$ channel of the ensemble a67 as an example to investigate the contributions from different diagrams D and C. First, we introduce
\beq\label{eq:c4_mass2}
\tilde{C}^{(4)}(t)&\equiv& C_{A_1}^{(4)}(t)-C_{A_1}^{(4)}(t+1) \nonumber \\
&=& d\left[ \cosh\left(E(t-\frac{T}{2})\right)-\cosh\left(E(t+1-\frac{T}{2})\right)\right] \nonumber \\
\eeq
where $E$ is the two-particle energy and $d$ is the coefficient. A shift of the four-point function can remove the constant terms. In real lattice simulations, the two-particle energy $E$ is obtained by solving the following equation
\beq
\frac{\tilde{C}^{(4)}(t)}{\tilde{C}^{(4)}(t+1)}=\frac{\left[ \cosh\left(E(t-\frac{T}{2})\right)-\cosh\left(E(t+1-\frac{T}{2})\right)\right]}{\left[ \cosh\left(E(t+1-\frac{T}{2})\right)-\cosh\left(E(t+2-\frac{T}{2})\right)\right]} \nonumber \\
\eeq
Then, the coefficient $d$ is extracted directly with the input of $E$ in Eq.~(\ref{eq:c4_mass2}). The results of $E$ and $d$ are shown in Fig.~\ref{fig:A1_mass2}. For each part of the diagram, an obvious plateau exists for a large enough $t$. We perform a correlated fit for these lattice data to a constant at a suitable time region. For the $A_1$ channel of ensemble a67, the time region is found to be $t/a=[27,31]$. The coefficient $d$ is determined by the input of $E$ and the value is chosen at the time slice $t/a=27$. The numerical values of $E$ and $d$ are summarized in Table~\ref{tab:aE_d}. 
\begin{table}[!htb]
\center
\caption{For $A_1$ channel of ensemble a67, numerical results of two-particle energies  $aE$ and coefficient $d$ defined in Eq.~(\ref{eq:c4_mass2}) for the diagram D, C, and D+C, respectively.}
\label{tab:aE_d}
\begin{ruledtabular}
\begin{tabular}{ccc}
  & $aE$ & $d$  \\
\hline
D+C& $2.0283(3)$   & $1.67(1)\times 10^{-23}$  \\
 D & $2.0274(3)$  & $1.98(2)\times 10^{-23}$    \\	
 C & $2.0227(3)$      & $-0.307(2)\times 10^{-23}$   \\
\end{tabular}
\end{ruledtabular}
\end{table}

Although the contribution of type (C) is much smaller than that of type (D), as indicated by the coefficient $d$, its non-zero value cannot be ignored. Moreover, since the two-particle energy of type (C) is also smaller, the contribution of which will become more important as $t$ increases. It suggests the nonzero coefficient $d$ and the lower two-particle energy of type (C) cause the contribution of this part to interfere with that of type (D), effectively increasing the total energy of the system and manifesting a repulsive interaction. The situations for other channels are similar, so we will not go into details here.

\bibliographystyle{apsrev4-2}

\bibliography{ref}

\end{document}